\documentclass[fleqn,usenatbib]{mnras}

\usepackage{newtxtext,newtxmath}

\usepackage[T1]{fontenc}

\DeclareRobustCommand{\VAN}[3]{#2}
\let\VANthebibliography\thebibliography
\def\thebibliography{\DeclareRobustCommand{\VAN}[3]{##3}\VANthebibliography}

\usepackage{graphicx}	
\usepackage{amsmath}	

\title[Cluster Masses from SDSS Multi-band Images]{Estimating Cluster Masses from SDSS Multi-band Images with Transfer Learning}

\author[Sheng-Chieh Lin et al.]{
Sheng-Chieh Lin,$^{1}$\thanks{E-mail: sli346@uky.edu}
Yuanyuan Su,$^{1}$
Gongbo Liang,$^{2}$
Yuanyuan Zhang,$^{3,4}$
Nathan Jacobs $^{5}$
and Yu Zhang$^{5}$
\\
$^{1}$Department of Physics and Astronomy, University of Kentucky, 505 Rose Street, Lexington, KY 40506, USA\\
$^{2}$Department of Computer Science, Eastern Kentucky University, 521 Lancaster Avenue, Richmond, KY 40475, USA\\
$^{3}$Fermi National Accelerator Laboratory, PO Box 500, Batavia, IL 60510, USA\\
$^{4}$Mitchell Institute for Fundamental Physics and Astronomy, and Department of Physics and Astronomy, Texas A\&M University, College Station, TX 77843-4242, USA\\
$^{5}$Department of Computer Science, University of Kentucky, 329 Rose Street, Lexington, KY 40506, USA
}

\date{Accepted XXX. Received YYY; in original form ZZZ}

\pubyear{2022}

\begin{document}
\label{firstpage}
\pagerange{\pageref{firstpage}--\pageref{lastpage}}
\maketitle

\begin{abstract}

The total masses of galaxy clusters characterize many aspects of astrophysics and the underlying cosmology.
It is crucial to obtain reliable and accurate mass estimates for numerous galaxy clusters over a wide range of redshifts and mass scales.
We present a transfer-learning approach to estimate cluster masses using the \textit{ugriz}-band images in the SDSS Data Release 12.
The target masses are derived from X-ray or SZ measurements that are only available for a small subset of the clusters.
We designed a semi-supervised deep learning model consisting of two convolutional neural networks.
In the first network, a feature extractor is trained to classify the SDSS photometric bands.
The second network takes the previously trained features as inputs to estimate their total masses.
The training and testing processes in this work depend purely on real observational data.
Our algorithm reaches a mean absolute error (MAE) of $0.232$ dex on average 
and $0.214$ dex for the best fold.
The performance is 
comparable to that given by redMaPPer, $0.192$ dex.  
We have further applied a joint integrated gradient and class activation mapping method to interpret such a two-step neural network. 
The performance of our algorithm is likely to improve as the size of training dataset increases.
This proof-of-concept experiment demonstrates the potential of deep learning in maximizing the scientific return of the current and future large cluster surveys.

\end{abstract}

\begin{keywords}
galaxies: clusters: general -- methods: data analysis -- techniques: image processing -- surveys
\end{keywords}



\section{Introduction}

    Clusters of galaxies, residing at the knots of the cosmic web filaments, 
    are the most massive gravitationally-bound systems in the Universe.
    The understanding of galaxy clusters is built upon
    the standard cosmology model and complicated baryon physics
    (e.g., \citealt{1989ApJS...70....1A}, \citealt{1996ApJ...462..563N}, \citealt{1998MNRAS.301..881E}, \citealt{2004ApJ...610..745L}, \citealt{2006ApJS..167....1B}, \citealt{2010MNRAS.406.1759M}).
    The abundance of massive halos across cosmic time is sensitive to the matter density, the normalization of the primordial perturbation, and the density of state.
    Halo mass functions constructed from samples of galaxy clusters provide a unique way to examine the cosmological models.
    In practice, the galaxy cluster mass estimates are subject to a variety of systematic effects particularly associated with the complicated substructures within a cluster (\citealt{2006MNRAS.369.2013R}).
    As a booming number of cluster cosmology studies emerge, it is paramount to obtain unbiased mass estimates through a careful handling of systematics.
    
    As multi-component systems, galaxy clusters are detectable at different parts of the electromagnetic spectrum.
    In optical and near infrared, cluster galaxies are luminous and can be effectively searched by ground-based observatories.
    The intra-cluster medium (ICM), on the other hand, can be observed in X-ray through the thermal Bremsstrahlung and from the spectral distortion of the cosmic microwave background (CMB) through the Sunyaev–Zeldovich effect (\citealt{1972CoASP...4..173S}).
    Multi-wavelength observations have been utilized to probe the total masses of galaxy clusters.
    An ideal mass estimate can be obtained through weak lensing, the distortion of the galaxy shape caused by the foreground gravitational potential field.
    Cluster masses can also be measured with the ICM pressure profile based on the assumption of hydrostatic equilibrium.
    Such hydrostatic mass estimates can be biased by non-thermal processes in the ICM related to gas motions, although {\sl Hitomi} observations reveal a remarkably quiescent atmosphere for the central region of the Perseus cluster (\citealt{2016}).
    The measurements of weak lensing mass or hydrostatic mass require deep and high quality observations that are yet achievable for a large number of galaxy clusters.
    It is thus practical to adopt mass proxies inferred from mass-observable relations for large cluster surveys.
    Common mass proxies are X-ray luminosity ($L_{\rm X}$, \citealt{2007ApJ...668..772M}), spectral temperature ($T_{\rm X}$), gas mass ($M_{\rm gas}$, \citealt{2016MNRAS.463.3582M}), and their product ($Y_{\rm X}=T_{\rm X} M_{\rm gas}$, \citealt{2006ApJ...650..128K}), the SZ flux ($Y_{\rm SZ}$, \citealt{2016A&A...594A..27P}), optical richness (\citealt{2014ApJ...785..104R}), and the velocity dispersion of member galaxies (\citealt{2015ApJ...799..214B}).
    Since $90\%$ of the baryon content in galaxy clusters is in the form of ICM (\citealt{2013A&A...555A..66L}, \citealt{2016A&A...592A..12E}),
    observables that trace the ICM, such as diffuse X-ray and SZ emissions, are relatively low-scatter mass proxies compared to optical observables such as cluster richness (\citealt{2012MNRAS.426.2046A}).
    
    The scatter in mass at a fixed mass proxy depends critically on the mass range and dynamical state of the cluster sample.
    Generally speaking, $M_{\rm gas}$ and $T_{\rm x}$ have a scatter of $10\% - 15\%$.
    $L_{\rm x}$ is the noisiest mass proxy from X-ray and requires a small number of photons to measure.
    It has a scatter of $20\% - 30\%$, while that of richness is $25\%$ (\citealt{2009ApJ...692.1033V}, \citealt{2010MNRAS.406.1773M}, \citealt{2015A&A...573A.118L}; \citealt{2014ApJ...783...80R}).
    The systematic uncertainties can be further reduced in these measurements when mass proxies are calibrated with weak-lensing measurements (e.g., \citealt{2016MNRAS.457.1522A}, \citealt{2020MNRAS.495..428C}).
    $Y_{\rm SZ}$ has been shown to be a robust mass proxy against different systematic uncertainties such as pressure profile variations and the asymmetric beam effects.
    It is also unaffected by cosmological dimming and allows us to probe high redshift clusters.
    The calibrated $Y_{\rm SZ}-M_{500}$ relation has a remarkably low scatter of $\sim 6\%$ (\citealt{2016A&A...594A..27P}).

    Observationally, X-ray and certain SZ signals are mostly accessible to space observatories, such as {\sl ROSAT}, {\sl XMM-Newton}, and {\sl Planck}, which are not comparable to ground-based observatories in terms of efficiency and costs.
    In general, optical-based mass proxies are more available and applicable for a larger number of clusters with a wider range of masses than ICM-based estimates.
    It is therefore essential to build a connection between the cluster masses derived from X-ray or SZ observables and their optical properties.
    We aim to achieve this goal using modern data-processing techniques.

    Machine learning methods have been widely applied in astronomy and astrophysics, e.g., to estimate the photometric redshift (\citealt{2016PASP..128j4502S}), to identify strong lensing signal (\citealt{2018MNRAS.473.3895L}), and to probe the core properties of galaxy clusters (\citealt{2020MNRAS.498.5620S}).
    In particular, Convolutional Neural Networks (CNN) have been shown to be a promising technique to study cluster masses.
    For example, \cite{2019ApJ...876...82N} used a CNN algorithm to predict cluster masses from synthetic Chandra X-ray images derived from the IllustrisTNG simulations.
    Meanwhile, several studies have been conducted to learn the dynamical cluster masses using the phase-space distribution of member galaxies.
    For example, \citealt{2019ApJ...887...25H} constructed two CNN models with different dimensions to infer cluster masses from the line-of-sight (LOS) velocities of member galaxies and their projected radial distance to the cluster center.
    \cite{2021MNRAS.501.4080K} adopted a 3D CNN on the 3D phase-space information (LOS and 2D projected spatial distribution) of member galaxies to estimate cluster masses.
    Moreover, to achieve robust estimates of cluster masses with uncertainties, various techniques have been employed in the training of CNN.
    \citealt{2020MNRAS.499.1985K} adopted a CNN with normalising flow mechanism to estimate the conditional probability distribution of cluster masses.
    \citealt{2021ApJ...908..204H} applied the variational Bayesian inference on the dynamical observables of member galaxies to estimate the cluster masses.
    However, to our best knowledge, previous machine learning studies of cluster masses all rely on simulated data since real observations often lack known ground truth and involve non-trivial noise and contamination.
    In this work, we present a novel transfer-learning approach in the framework of CNN to estimate ICM-based mass proxies using the photometric images from the Sloan Digital Sky Survey (SDSS, \citealt{2017AJ....154...28B}).

    With the on-going and upcoming large-area surveys, such as the Dark Energy Surv ey (DES, \citealt{2018ApJS..239...18A}) and the Vera C. Rubin Observatory (previously known as LSST), machine learning techniques are expected to play an important role in enhancing the scientific returns of the tremendous amount of in-flowing data.
    Our pioneering study using purely real observations can pave the way for this promising future.
    This paper is structured as follows.
    We describe the data preparation of the images and catalogues in Section \ref{sec:data}.
    Section \ref{sec:methods} is devoted to the model design and training strategy.
    We present the result and compare it with the performance of traditional methods in Section \ref{sec:results}.
    We interpret the models in Section \ref{sec:discussion}, and discuss our findings in Section \ref{sec:summary}.
    Throughout this work we adopt a flat $\Lambda$-CDM cosmology from \cite{2016A&A...594A..13P}:
    $\Omega_{m} = 0.3089$, $\Omega_{\Lambda}=0.691$, $\Omega_{b}=0.0486$, $H_0=67.74\, \text{km}\text{s}^{-1}\text{Mpc}^{-1}$.

\section{Data Preparation}
\label{sec:data}

    We aim to directly predict the ICM-based mass proxy for a cluster from its multi-band photometric images.
    In this section, we describe the construction of our samples.
    
    \subsection{SDSS images and redMaPPer catalogue}

        To acquire a sizable cluster sample from a single survey, we use the $ugriz$-band image data from SDSS Data Release 12 (DR12, \citealt{2015ApJS..219...12A}).
        The survey covers roughly $31000$ deg$^2$ of the sky with the depth of $r$-band magnitude to $22.2$, providing a large sample of galaxy clusters.
        The broad-band images used in training are background subtracted and calibrated by the SDSS imaging pipeline (\citealt{2001ASPC..238..269L}; for an updated version, \citealt{2011ApJS..193...29A}).
        
        To select a parent sample of galaxy clusters, we make use of the cluster catalogue constructed by the red-sequence Matched-filter Probabilistic Percolation (redMaPPer) algorithm using SDSS DR8 data (\citealt{2014ApJ...785..104R}).
        In short, the algorithm detects galaxy clusters using the colours and the spatial distribution of the red sequence galaxies.
        The probability of being a member of a cluster is assigned to each individual galaxy.
        The sum of these probabilities within a given radius, called richness ($\lambda$), approximates the number of red-sequence galaxies within a cluster.
        Richness is therefore correlated with the optical content of a cluster and can be treated as an optical-based mass proxy for galaxy clusters.
        Based on the catalogue, we have selected $\sim 25000$ clusters with richness greater than $20$.
        
        The SDSS photometric images of the clusters are obtained by applying the following searching criteria.
        For each cluster, we select all the images within a square with a length of $4\,\text{Mpc}$ centred at the cluster centre listed in the catalogue.
        These images are further mosaic-ed using an image processing package, SWarp (\citealt{2002ASPC..281..228B}).
        When mosaicing, the images are co-added by the median pixel value, and re-sampled using the \texttt{LANCZOS3} filter.
        The resulting images are central-cropped to a physical size of $2\,\text{Mpc}$.
    
        \subsection{Image pre-processing}
    \label{subsec:pre}
        
        A global constant was added to all the input images to avoid negative pixels in the images due to the background subtraction adopted by the SDSS imaging pipeline.
        This treatment may introduce a bias to the magnitude measurements of individual pixels.
        However, it does not affect the relative differences among pixels and images on which our model focuses.
        The pixel values are converted to a log scale and normalized to the range of 0 and 1.
        All the input images over a physical size of $2\times2\,\text{Mpc}^2$ centred on each cluster are resized to have a dimension of $256\,\text{pixel}\,\times\,256\,\text{pixel}$.
        Each single band image is triply-stacked to produce a 3-channel image with identical channels to match the number of input channels used by the network.
        The input images are augmented during the training process by a random combination of a horizontal/vertical flip and/or a $0/90/180/270$ degree rotation.
            
    \subsection{Mass estimates}
    
        We use $M_{500}$ to characterize the total mass of each cluster, defined as the enclosed mass within $R_{500}$, a radius within which the average density is 500 times the critical density of the Universe at that redshift.
        It is desirable to obtain ICM-based mass estimates for as many clusters as possible, whereas the availability of their SZ or X-ray measurements is limited.
        We make use of the SZ source catalogue from the {\sl Planck} mission 2015 data release for SZ detected clusters.
        The MCXC, XCS, and RASS-DR7 catalogues are used for X-ray detected clusters that are not included in the {\sl Planck} SZ catalogue already.
        
        \textit{Planck}:
        The cluster catalogue built from the {\sl Planck} 2015 data release (\citealt{2016A&A...594A..27P}) contains $1653$ clusters with redshifts spanning from $0.01$ to $0.9$.
        Three algorithms have been implemented to detect clusters in the CMB map: two of them are based on the matched multi-filtering technique and the third one uses Bayesian inference.
        The cluster mass is estimated from the Compton $Y$-parameter using the $Y_{500}-M_{500}$ relation presented in \cite{2014A&A...568A..22B}.
        The redshifts of these clusters are determined through the counterparts matched in external data sets, including MCXC, SDSS redMaPPer catalogue, ACT, and SPT cluster catalogue.
        
        \textit{MCXC}:
        The Meta-catalogue of the compiled properties of X-ray detected Clusters of galaxies (MCXC) is composed from nine publicly available {\sl ROSAT} All-sky survey-related cluster catalogues and seven supplemental cluster catalogues from other {\sl ROSAT} surveys, such as The 160 Square Degree {\sl ROSAT} Survey.
        The data has been homogenised in terms of coordinates, names, soft X-ray luminosities, and masses.
        Duplicated clusters have been removed.
        The cluster masses are derived using the $L_{\rm X}-M_{500}$ relation (\citealt{2010A&A...517A..92A}).
        \cite{2011A&A...534A.109P} provides more details on the construction of the catalogue.
        
        \textit{XCS}:
        The {\sl XMM} Cluster Survey (XCS) catalogue is constructed from the publicly available database, XMM-Newton Science Archive, for the purpose of studying cosmology and the evolution of cluster X-ray properties.
        It contains $503$ optically confirmed X-ray clusters with redshift ranging from $0.06$ to $1.46$.
        The detailed construction of this catalogue is given by \cite{2012MNRAS.423.1024M}.
        We derive the cluster masses directly from the cluster radius $R_{500}$ provided in the catalogue.
        
        \textit{RASS-DR7}:
        This catalogue is constructed by \cite{2014MNRAS.439..611W} based on the SDSS galaxy group catalogue of \cite{2007ApJ...671..153Y}.
        The X-ray properties of these optically-selected clusters are taken from the {\sl ROSAT} broad-band X-ray images.
        We select clusters with SNR $>1$ that are not included in other X-ray or SZ catalogues.
        Their masses are estimated using the $L_X-M_{500}$ scaling relation calibrated for galaxy groups from \citealt{2010MNRAS.406.1773M}.
        
        We cross-match the redMaPPer catalogue with the above X-ray and SZ cluster catalogues.
        We require the difference of the cluster redshifts to be within $\delta(z) < 0.05$ and the projected separation of the cluster centres to be within $10$ arcmin.
        Spectroscopic redshift is used if available in these catalogues; otherwise, photometric redshift is adopted.
        To avoid assigning multiple X-ray/SZ sources to the same cluster, we further impose that the 3D spatial separation between the cluster centres is within $500$ kpc.
        The matching criteria applied here are slightly stricter than that applied in the cylindrical matching in \cite{2014ApJ...783...80R}.
        A similar projected matching criteria can also be found in \cite{2014A&A...571A..87S}.
        The cross-matching criteria are summarized below:
        \begin{enumerate}
            \item $\delta(z) < 0.05 $
            \item $\delta(\theta) < 10^{\prime}$
            \item $\delta(r) < 500\, \text{kpc}$
        \end{enumerate}
        
\begin{figure}
            \centering
            \includegraphics[width=\columnwidth]{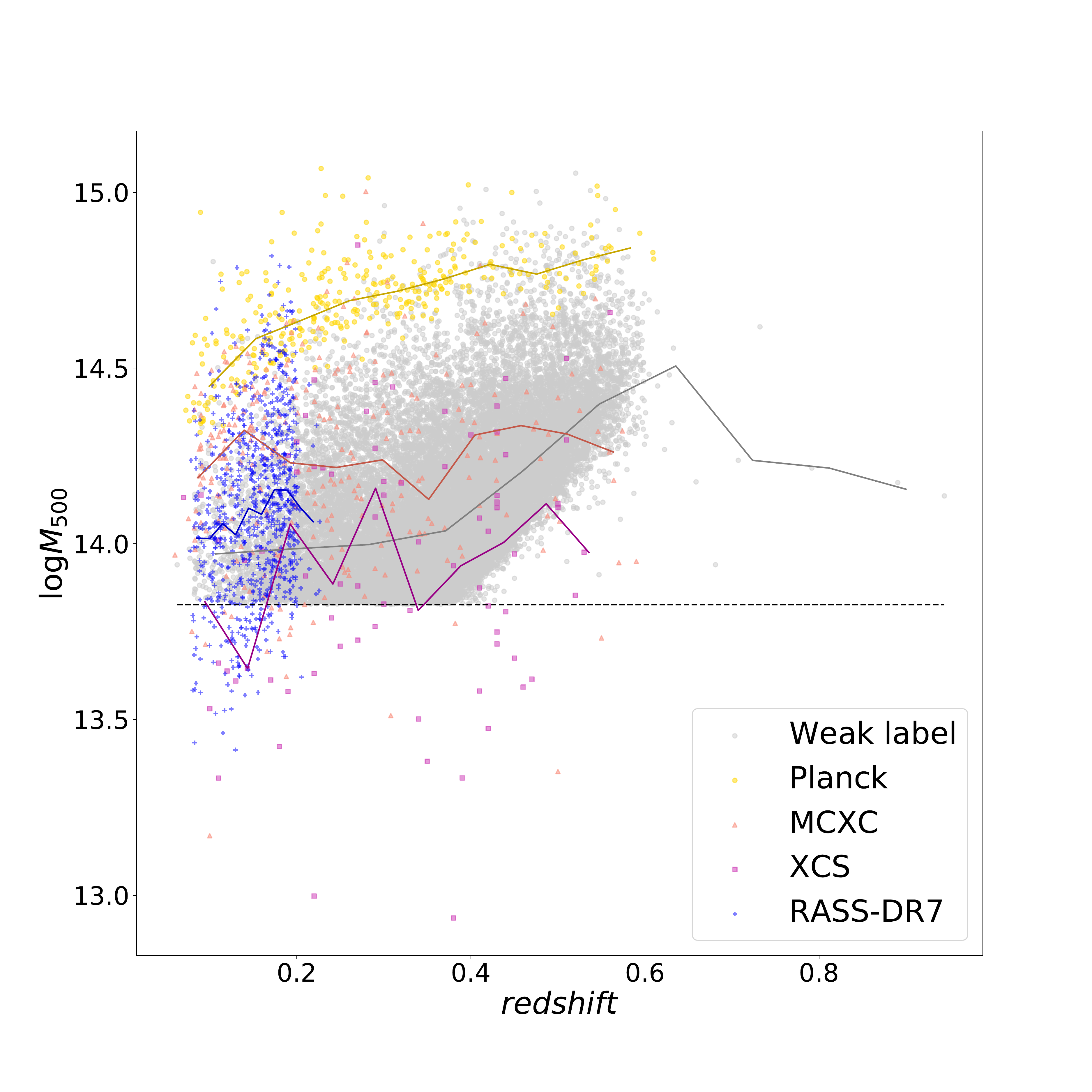}
            \caption{
                Cluster mass as a function of redshift for clusters in our sample.
                The distributions of different source catalogues are colour-coded separately.
                Those without ICM-based mass estimates are denoted in grey.
                The solid lines represent the median of $\log M_{500}$ in each redshift bin width of $\delta_z = 0.01$.
                The dashed line shows the richness cut $\lambda=20$ applied in the redMaPPer catalogue.}
            \label{fig:mass_as_fn_z}
        \end{figure}

        The cluster sample is divided into two subsets, strong and weak labels, based on the availability of their X-ray/SZ counterparts.
        For strong labels, we obtain $1515$ SDSS redMaPPer clusters for which the ICM-based masses are available.
        Of these clusters, $340$ are from the {\sl Planck} catalogue, $271$ are from  MCXC, $78$ are from XCS, and $826$ are from RASS-DR7.
        Cluster properties other than the halo mass estimates such as centre, redshift, richness are taken from the redMaPPer catalogue.
        The mass-redshift distribution of all the clusters in our sample is shown in Fig. \ref{fig:mass_as_fn_z}.
       
        The rest of the $24611$ redMaPPer clusters without X-ray or SZ counterparts are used as weak labels with the cluster masses derived from the richness-mass relation given by \citealt{2018MNRAS.478..638J}, which takes the form of a powerlaw:
        \begin{equation}
            \langle M_{500} \rangle (\lambda) = M_0 \left(\dfrac{\lambda}{\lambda_0}\right)^{\alpha_{M|\lambda}}
            \label{eq:mass-richness}
        \end{equation}
        where $M_0$ is the reference mass defined at $\lambda=\lambda_0$.
        We follow the best-fit values given in the paper where $M_0 = 10^{14.434}\,M_{\odot}$, $\alpha=1.23$, and a fixed $\lambda_0 = 60$.
        These values are obtained by comparing the mass bias measurement and two richness-mass relations obtained from the weak-lensing analysis in \cite{2017MNRAS.469.4899M} and \cite{2017MNRAS.466.3103S}.
        
\section{Methods}
\label{sec:methods}

    To build the mass prediction model, we adopt a two-step semi-supervised approach to utilize both the large weak label sample and the small strong label sample.
    Conventionally, semi-supervised learning refers to a category of machine learning using both unlabeled and labeled data to achieve better predictions than those solely based on labeled data (\citealt{zhu2009introduction}, \citealt{van2020survey}).
    The need of this type of learning is due to lack of sufficient labeled data to train a deep network.
    Our model falls into this category because a larger set of weak-labeled data (although it is not unlabeled) has been used to facilitate the training.

    In the first step, we perform a $5$-class classification task to predict the photometric band of each image in the combined sample including both weak and strong label data.
    In the second step, we perform a regression task to estimate the ICM-based mass, $M_{500}$, for the strong label data.
    The neural networks of these two tasks are connected through transfer learning.
    These networks are implemented in PyTorch (\citealt{NEURIPS2019_9015}).
    The model architecture and the training procedure are described in detail in the following subsections.

    \subsection{Neural network architecture}
    
        \begin{figure*}
            \centering
            \includegraphics[width=2.\columnwidth]{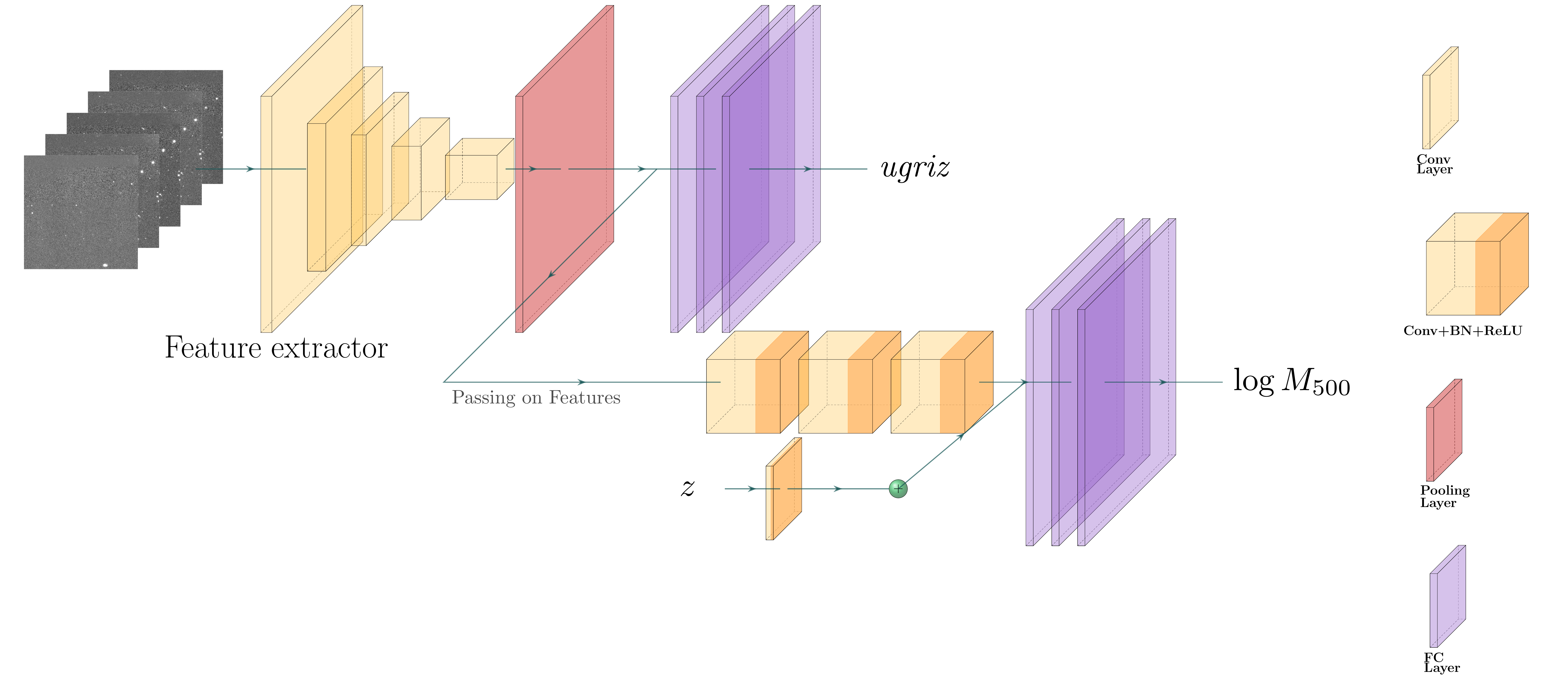}
            \caption{The full architecture of our model.
                     The feature extractor is solely based on the ResNet-18 architecture.
                     After the GAP layer, we divide the outputs into two branches: one is directly connected to three FC layers which outputs the photometric bands and the other branch leads to the next training of the mass estimation.
                     The green plus sign after the 1D Conv layer indicates the concatenation of the features from both types of the Conv layers in the regression network.}
            \label{fig:network}
        \end{figure*}

        \begin{table}
        	\centering
        	\caption{Layer parameters of the regression network. The input size of 2D Conv 1 is $1 \times 512 \times 5$, which is the size of the stacked feature vectors from the feature extractor. The outputs from the last 2D Conv layer and the 1D Conv layer are flattened and concatenated to be fed into FC 1 layer.}
        	\begin{tabular}{lccr} 
        		\hline
        		Layer & kernel & output size\\
        		\hline
        		Input: feature maps \\
        		2D Conv 1 & $1\times 1$ & $1\times 512 \times 5$ \\
        		BN & $-$ & $1\times 512 \times 5$ \\
        		ReLU & $-$ & $1\times 512 \times 5$ \\
        		2D Conv 2 & $3\times 3$ & $16\times 510 \times 3$ \\
        		BN & $-$ & $16\times 510 \times 3$ \\
        		ReLU & $-$ & $16\times 510 \times 3$ \\
        		2D Conv 3 & $3\times 3$ & $64\times 508 \times 1$ \\
        		BN & $-$ & $64\times 508 \times 1$ \\
        		ReLU & $-$ & $64\times 508 \times 1$ \\
        		\hline
        		Input: catalog \\
        		1D Conv & $1\times 1$ & $1\times 1 \times 1$ \\
        		BN & $-$ & $1\times 1 \times 1$ \\
        		ReLU & $-$ & $1\times 1 \times 1$ \\
        		\hline
        		FC 1 & $-$ & $1024$ \\
        		FC 2 & $-$ & $512$ \\
        		FC 3 & $-$ & $1$ \\
        		\hline
        	\end{tabular}
        	\label{tab:layer_info}
        \end{table}

        We use a residual network (ResNet; \citealt{2015arXiv151203385H}), in particular the well-known ResNet-18 architecture, as the core feature extraction architecture.
        Residual networks are a class of CNNs that use residual connections to facilitate faster training.
        Such networks have been widely used for image classification applications. Like other CNNs, they are composed of sets of trainable convolutional filters, which learn to extract features based on the task the network is being optimized for.
        Such networks can learn to detect low-level features, such as curves and edges, and higher-level features, such as human faces and vehicles.
        The convolutional filters are arranged roughly in sequence, and are interspersed with pooling layers, which are used to reduce the spatial resolution, batch normalization (BN) layers, and nonlinear activation layers (e.g., the rectified linear unit), which enable the learning of nonlinear mappings.
        Two types of pooling are used: max pooling and average pooling, which respectively compute the maximum value and the average value in local windows across the image.
        These two pooling operations can also be applied to the entire image, called global max pooling and global average pooling (GAP).
        The GAP layer is typically near the end of the network, and marks the transition from extracting spatial features to computing properties of the full image. In addition, it provides flexibility by allowing images of different sizes.
        
        Neural network-based approaches are often considered uninterpretable, {\em black box} methods. However, there is often useful structure that can be observed in the extracted features. The convolutional layers that are closer to the input image typically extract the low-level features and the layers closer to the output extract the high-level features, which are based on the low-level features. In addition, it can be useful to view the network as being composed of two major modules: one is a linear classifier or regression model, which is just the final layer of the network, and the other is a feature extraction module that is optimized to translate raw inputs into a suitable feature representation for the linear model.
        
        The residual connection is the unique architecture component that was popularized by the introduction of residual networks. It is realised by adding the input of a module $f_i$, which can consist of several convolutional and other layers, to the output of the module as follows:
        \begin{equation}
             y_i = \phi(f_i(x_i;\Theta_i)+x_i),
        \end{equation}
        where $x_i$ is the input, $f_i(\cdot)$ represents the mapping with trainable parameters $\Theta_i$, $\phi(\cdot)$ is the non-linear activation function, and $y_i$ is the output.
        
        The full architecture of our model is shown in Fig.\ref{fig:network}.
        In the band classification task, ResNet-18 is used as the feature extractor.
        All layers before the GAP layer in the ResNet-18 are used and followed by an GAP layer, and three fully connected (FC) layers.
        The network is initialized with  ImageNet weights, which have been shown to be widely useful for transfer learning.
        Note that, as mentioned in Sec.\ref{subsec:pre}, we stack a single band image into a 3-channel image to avoid changing the original ResNet-18 architecture and to allow us to initialize the network with the ImageNet weights.
        As for the regression task, we use a relatively simple architecture consisting of three 2D Conv layers, a 1D Conv layer, and three FC layers at the end.
        A BN layer and a ReLU layer are connected after each of the Conv layers.
        The first 2D Conv layer takes the feature vectors from the classification network as the inputs.
        The 1D Conv layer is used to include the information on redshift $z$, and its output is directly concatenated at the end of the output vectors of the 2D Conv layer, together to be fed to the FC layers.
        The layer information of the regression network is shown in Table \ref{tab:layer_info}.
        The number of parameters of the feature extractor is $\sim 1.25\times10^7$ and that of the regression network is $\sim 3.38\times10^7$.
            
        We note that while we used the ResNet-18 architecture, there are many alternatives that could be used.
        We selected it by empirical evaluation over alternatives, including VGG-16 (\citealt{2014arXiv1409.1556S}), GoogLeNet (\citealt{2014arXiv1409.4842S}), and ResNet-50, which is a deeper variant of ResNet-18.

    \subsection{Training procedure}
        
        The semi-supervised transfer learning model consists of two CNNs which perform the classification task on SDSS photometric bands and the regression task on estimating masses, respectively.
        The band classification training is performed on the feature extractor, which outputs the predicted photometric bands and the feature vectors for later use.
        The learning procedure is carried out by minimizing the cross-entropy loss function.
        The performance of feature extractor is evaluated by employing a train-test split with $80\%$ of data used for training and the rest for testing.
        The training iterates for $100$ epochs and reaches a convergence within $50$ epoch.
        We run the training for longer epoch to make sure that over-fitting does not occur.
        As a result, the best accuracy of band predictions on test set is $92\%$.
        This high accuracy implicitly indicates the feature extractor has built up knowledge about the photometric images.

        For the regression training, the second network takes the feature vectors obtained from the GAP layer of the extractor as the input to be trained to predict the cluster masses.
        As a pretraining strategy,  we first train the regression network on weak label data.
        Initialized with the pretrained weights, the network is further trained on strong label data to produce the final results.
        We also feed the network with the cluster redshift $z$ using a 1D Conv layer to help break the degeneracy between cluster sizes and distances.
        Redshift is provided with a trainable parameter that the network by itself can decide to take it into account or not.
        The loss function used in the regression training is defined as follows,
        \begin{equation}
            \label{eq:loss}
            \textit{L} = \dfrac{1}{N} \sum_{i=1}^{N} w_i \left( y_i - \hat{y}_i \right)^2 + 100 \sigma_{\Delta},
        \end{equation}
        where $y_i$ is the output of the model, $\hat{y}_i$ is the ground truth, $\sigma_{\Delta}$ denotes the $1\sigma$ standard deviation of the difference between $y_i$ and $\hat{y}_i$, and $w_i$ is the weight accounting for the imbalance of the data, which is inversely proportional to the number of clusters in a mass range.
        This loss is specifically designed to encourage the network to minimize the difference between the ground truth and the predictions, and to constrain the distribution of the predictions.
        For the latter term in the loss function, the scalar multiplied on $\sigma_{\Delta}$ is empirically chosen.
        The Adam optimizer (\citealt{2014arXiv1412.6980K}) with a learning rate of $L_{r} = 0.0001$ is adopted to update the network for $100$ epochs, which is sufficient for the training to converge.
        Additionally, a $10$-fold cross-validation is used to guarantee the generality of the model.
        In short, the strong label data is evenly divided into $10$ subsets with no repetition.
        Eight subsets are used as the training set, one subset as the validation set and one as the testing set for each fold.
        This process is repeated $10$ times to cycle through all the data.

\section{Results}
\label{sec:results}

\begin{figure}
            \centering
            \includegraphics[width=\columnwidth]{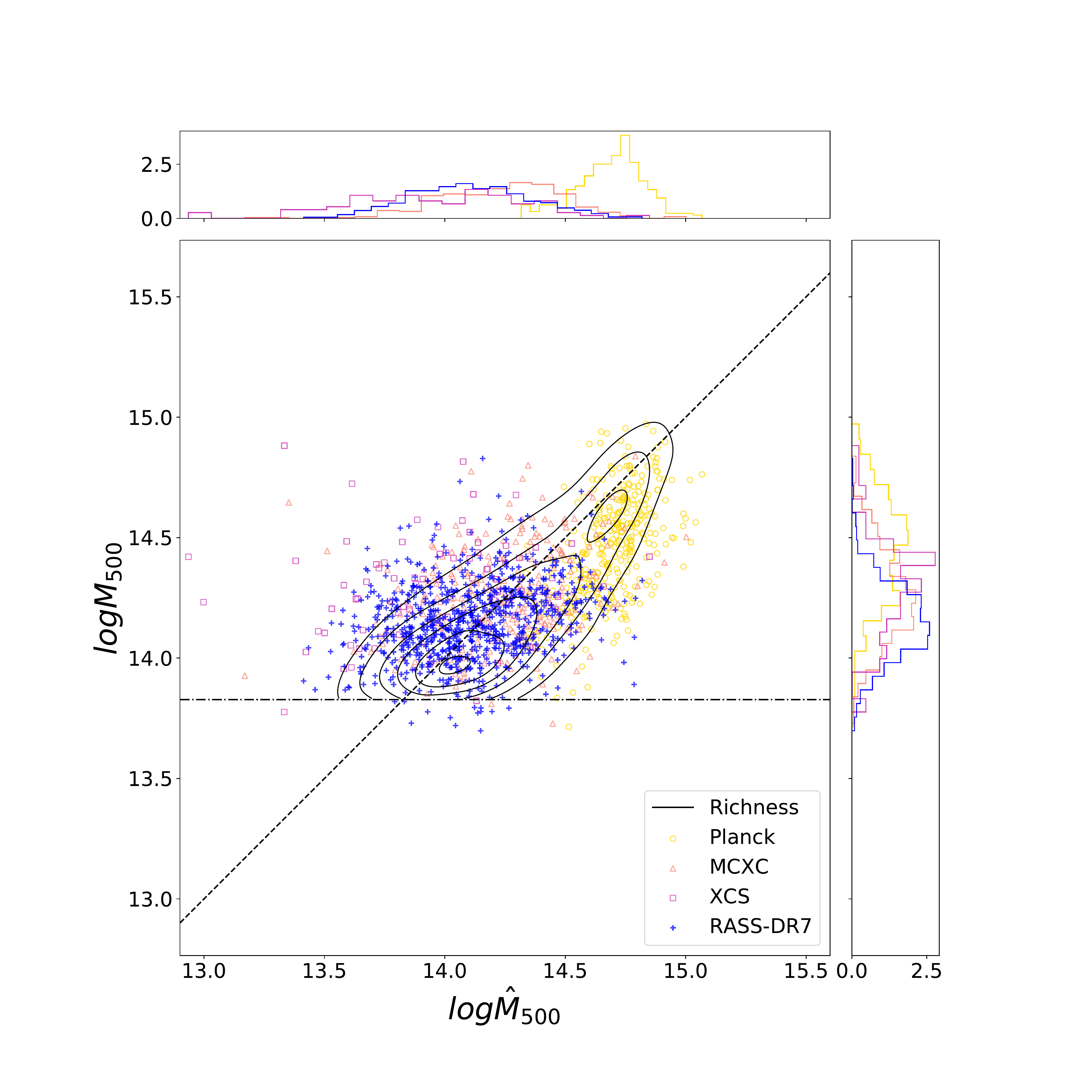}
            \caption{
                Comparison between the predicted masses $\log M_{500}$ and the target values $\log \hat{M}_{500}$.
                Four sets of masses obtained from different source catalogues are colour-coded separately.
                The richness-based mass as a function of target value is shown as the coloured contour.
                The horizontal dashed-dotted line shows the richness cut of $20$ applied in the original redMaPPer catalogue.
                The probability distributions of predicted and target masses of four subsets (weak label excluded) are shown in the above and right panels, respectively.}
            \label{fig:main_result}
        \end{figure}

    We compare the predicted cluster masses and their target values in Fig.\ref{fig:main_result}.
    The following statistics are measured to evaluate the performance of the model: the mean absolute error (MAE) is defined as
    \begin{equation}
        \label{eq:mae}
        \text{MAE} = \dfrac{1}{N} \sum_{i=0}^{N} \left\vert \log{M_{500}} - \log{\hat{M}_{500}} \right\vert,
    \end{equation}
    where $\log{M_{500}}$ is the predicted mass, $\log{\hat{M}_{500}}$ is the target ICM-based mass, and $N$ is the batch size;
    the standard deviation $\sigma_{\Delta \log M}$ of the mass residual in log scale, which is defined as
    \begin{equation}
        \label{eq:diff_m}
        \Delta \log M = \log{M_{500}} - \log{\hat{M}_{500}}.
    \end{equation}
    The best fold has an MAE of $0.214$ dex.
    The MAE averaged over $10$ cross-validation folds is $0.232$ dex.
    The cross-validation standard deviation of MAE is $0.012$ dex showing a stable result across $10$ folds.
    The standard deviation of the residual averaged over $10$ folds is $0.292$ dex.
        
    \subsection{Compared with richness-based mass estimates}
    
        For comparison, we also derive the expected cluster mass for each cluster based on the mass-richness relation shown in Eq.\ref{eq:mass-richness}.
        The redMaPPer algorithm applied to the SDSS catalogue data reaches an MAE of $0.192$ dex and a standard deviation of the residual $\sigma_{\Delta \log M} = 0.249$ dex, which is superior but comparable to our model directly applied to SDSS photometric images.

\begin{figure}
            \centering
            \includegraphics[width=\columnwidth]{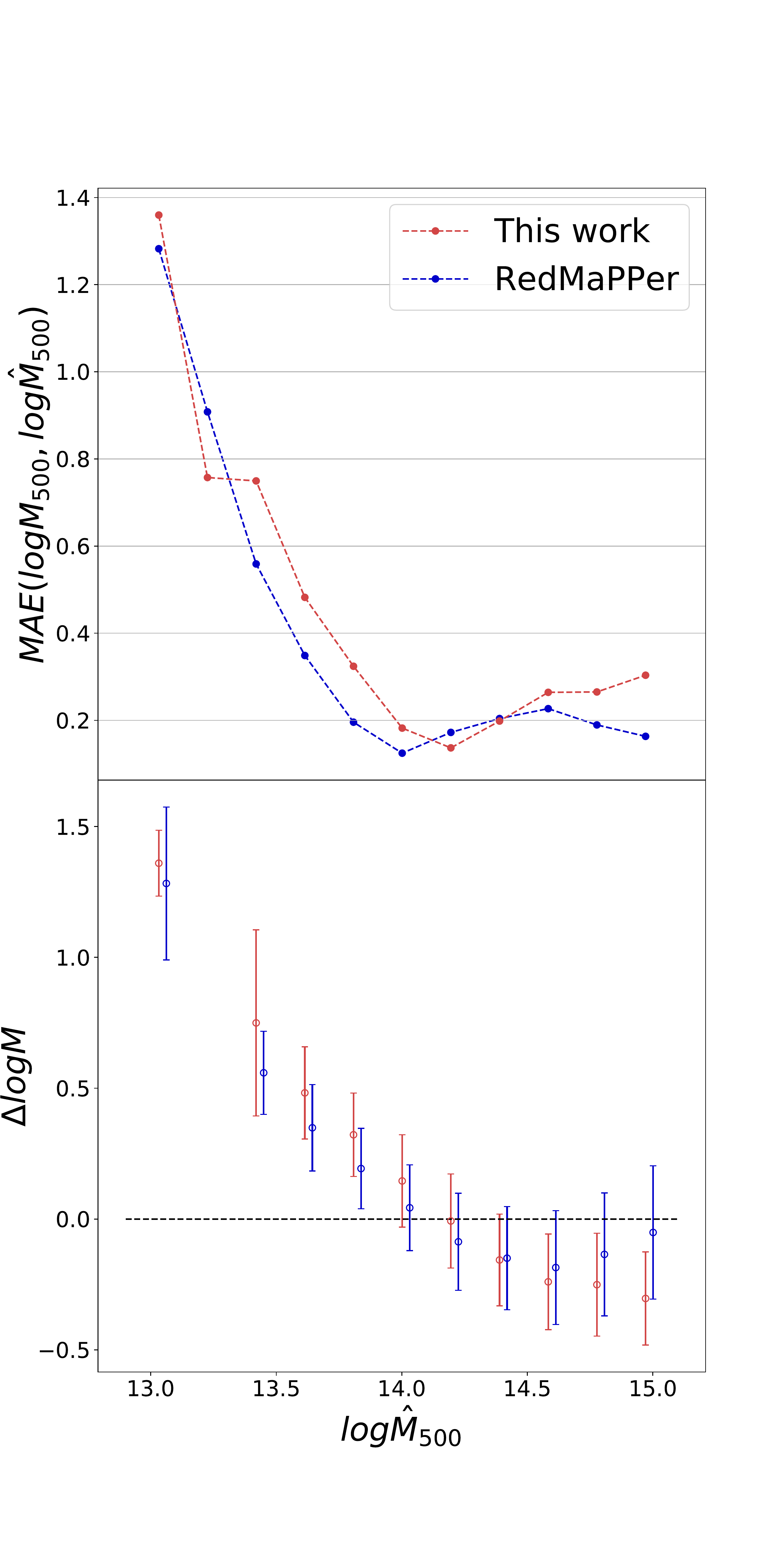}
            \caption{The performance of our deep learning approach in comparison with that of redMaPPer as a function of cluster mass scale.
                     \textit{Upper panel}: the comparison of the MAE defined in Eq.\ref{eq:mae}.
                     \textit{Lower panel}: the comparison of the residue defined in Eq.\ref{eq:diff_m}.
                     The errorbar denotes the $1\sigma$ error.
                     The redMaPPer result is slightly shifted to the right for clarity.
                    }
            \label{fig:diff_at_massese}
        \end{figure}

        To understand the dependence of the performance on cluster mass scales, we compare the values of MAE and  $\sigma_{\Delta \log M}$ as a function of the target value with those derived from redMaPPer.
        As shown in Fig.\ref{fig:diff_at_massese}, the performance of our model in terms of both MAE and the scattered residual is comparable to that of redMaPPer at all mass scales.
        Both our model and redMaPPer estimate the masses accurately for galaxy clusters with masses above $10^{14}\,M_{\odot}$.
        The accurate estimates are likely due to a higher purity, the ratio of the member galaxies to that of the observed galaxies within a given FOV, for massive clusters, such that the effects of interlopers and background noise are reduced.
        On the other hand, a clear overestimating trend at the low-mass end can be seen in the lower panel of Fig.\ref{fig:diff_at_massese} for both methods.
        The overestimate in redMaPPer sample is mainly caused by the richness cut at $N=20$ which removes potential low-mass clusters.
        In our case, the characterization of low-mass clusters with fewer member galaxies are more vulnerable to interlopers and background noise.
        We note a slightly worse performance of our model at the very high-mass end (above $10^{14.5}\,M_{\odot}$) which is not observed in redMaPPer.
        This may be related to the mean-reversion edge effect as the model prediction for the end of the target value range can be biased towards the mean (\citealt{2021ApJ...908..204H}; \citealt{2021MNRAS.501.4080K}).
        This bias might be mitigated once we obtain sufficient number of clusters above $10^{14.5}\,M_{\odot}$.
        In general, it is desirable to have more labeled data at both ends to improve the performance even though our approach has already greatly reduced the need of strong label data.
        
\begin{figure}
            \centering
            \includegraphics[width=\columnwidth]{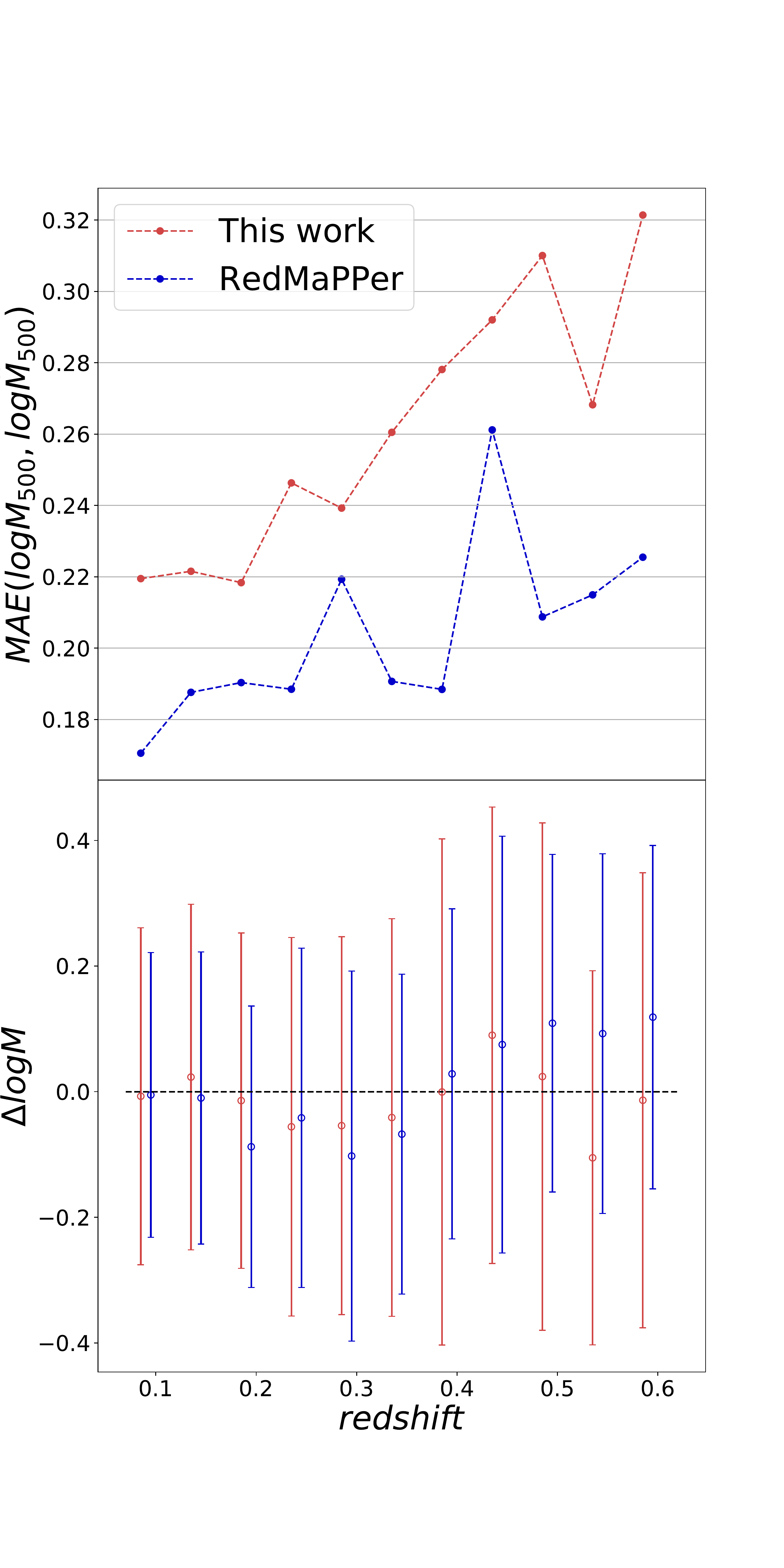}
            \caption{The dependence of performance on redshift for our deep learning approach and the redMaPPer algorithm.
                     The format is the same as Fig.\ref{fig:diff_at_massese}.
                     \textit{Upper panel}: the MAE values as a function of redshift.
                     \textit{Lower panel}: the mass residual as a function of redshift with $1\sigma$ errorbars.
                    }
            \label{fig:diff_at_size_and_z}
        \end{figure}

        Cluster redshift is another important factor since it affects the apparent size and SNR of a cluster.
        In Fig.\ref{fig:diff_at_size_and_z}, we show the dependence of the predicted mass on redshift and compare it with redMaPPer.
        For both methods, a gradually increasing trend of MAEs toward higher redshift is observed.
        The performance of our model has a higher MAE than that of redMaPPer at all redshifts 
        as shown in the lower panel of Fig.\ref{fig:diff_at_size_and_z}.
        
    \subsection{Compared with plain ResNet-18 model}
    
        In this subsection, we demonstrate the necessity of the transfer learning approach for the task of estimating cluster masses from optical images.
        We compare the performance of our semi-supervised transfer learning model (S.S.+Trans.) with a baseline model using the ResNet-18 architecture which is trained on 5-channel (photometric bands are treated as channels) $256\times256$ images to estimate the ICM-based masses.
        We apply the same training procedure and regression loss function given in Sec.\ref{sec:methods}; however, we do not use the ImageNet weights to initialize the baseline model due to the modification of the input layer.
        To show the improvement given by the semi-supervised transfer learning, we set up two experiments: (1) training baseline on strong label data and (2) semi-supervised training (baseline+S.S.) on weak label data and strong label data.
        
\begin{figure}
            \centering
            \includegraphics[width=\columnwidth]{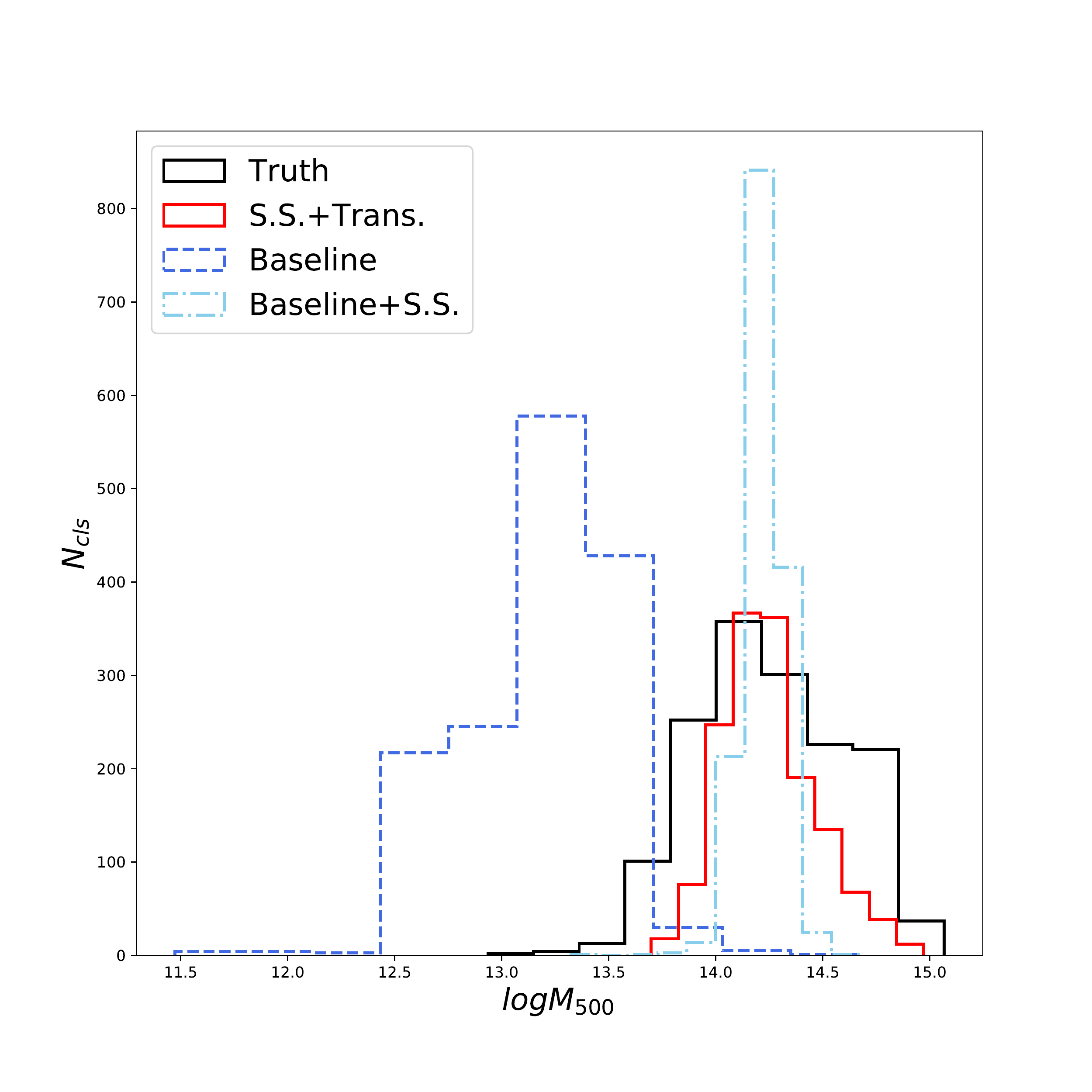}
            \caption{The distribution of predicted masses from several different models and the target values (ICM-based masses).
                     Here we show the results from ResNet-18 baseline (Baseline), ResNet-18 baseline with semi-supervised learning (Baseline+S.S.), and our approach (S.S.+Trans.).
                     The target value, $\log \hat{M}_{500}$, is denoted in black.
                     Both baseline models are shown in the dashed lines and the red solid line is our approach.}
            \label{fig:comp_w_baselines}
        \end{figure}

        Table \ref{tab:comp_w_baselines} shows the cross-validation MAEs of three models: baseline, baseline+S.S., and our approach, S.S.+Trans.
        In Fig.\ref{fig:comp_w_baselines}, we compare the distributions of ground truth as well as the results from these three models.
        As expected, the baseline model trained directly on strong label data gives the worst performance since the amount of data is extremely insufficient for a deep network like ResNet-18 to learn meaningful features.
        We also find that in Fig.\ref{fig:comp_w_baselines} the baseline+S.S. results are distributed sharply around the median of $\log M_{500}=14.218$.
        It shows that the model does not acquire enough information about masses, even though a much larger dataset with $\sim 25000$ clusters is used in the pretraining.
        By comparing these three models, we conclude that the S.S.+Trans. model learns better from the data and performs well on the given task.

        \begin{table}
        	\centering
        	\caption{Cross-validation MAEs of the two baselines and our model.}
        	\begin{tabular}{lccr} 
        		\hline
        		Models & cv-MAE Ave. & cv-MAE $1\sigma$\\
        		\hline
        		Baseline & 1.069 & 0.302 \\
        		Baseline+S.S. & 0.28 & 0.018 \\
        		S.S.+Trans. & 0.232 & 0.012 \\
        		\hline
        	\end{tabular}
        	\label{tab:comp_w_baselines}
        \end{table}
        
\section{Model Interpretation}
\label{sec:discussion}
    
    To interpret the mechanism, we utilize the integrated gradients (IG, \citealt{2017arXiv170301365S}) method and the Class Activation Mapping (CAM, \citealt{2015arXiv151204150Z}) method to quantify the importance of the input features.
    The two-step transfer learning approach involves two networks -- the feature extractor and the regression network, which take different forms of inputs and outputs.
    We therefore divide the model interpretation into two parts accordingly.
    
    The interpretation method used in both networks is IG, which attributes the output of a network to the input features by integrating the path along the gradients from a baseline input to the original input.
    The choice of baseline can be vastly different with distinct impacts on the interpretation (\citealt{sturmfels2020visualizing}).
    The formal definition of IG is as follows:
    \begin{equation}
        IG_i \Coloneqq \left(x_i - x_i^{b} \right)
            \int_{0}^1 \nabla F\left(\alpha x_i + (1-\alpha)x_i^b \right) d \alpha
        \label{eq:ig}
    \end{equation}
    where $\nabla F(\cdot)$ is the gradients of the model, $x_i$ is the $i$-th feature of the original input, and $x_i^b$ is the baseline at the same spatial location of $x_i$.
    This method has an advantage that it overcomes the issue of the saturate gradient problem commonly occurred in the gradient-based attribution methods.
    In general, one should avoid using a zero baseline if zero represents certain features of the inputs as suggested by \cite{sturmfels2020visualizing}.
    In our case, we use a zero baseline, which is a set of black images, because the re-scaled input images do not contain pixels with the value of zero across five bands.
    Additionally, the zero baseline allows us to examine whether the network focuses on the background noise or not.

    CAM, the other interpretation method used for the regression network, gives the highest ranked features by weighted-summing the last Conv layer at different spatial locations:
    \begin{equation}
        M(x,y)=\sum_{k} w_k f_k(x,y),
    \end{equation}
    where the subscript $k$ denotes a neuron unit in the last Conv layer, $w_k$ is the weight of that neuron, and $f_k(x,y)$ is the activation function of that neuron at spatial position $(x,y)$ in a given image.
    In principle, the weight $w_k$ indicates the importance of such feature with respect to a given class for the classification training.
    Given the design of our model, we adopt a slightly different version where the IG maps are treated as weights instead of the class weights in traditional CAM.
    
    \subsection{Feature extractor}
    
        For the feature extractor, the interpretation is based solely on the IG method (Eq.\ref{eq:ig}).
        In principle, brighter objects in images should gain more attention from the network since they are easier to capture.
        A set of examples obtained by IG are shown in the middle row of Fig.\ref{fig:ig}.
        We find different patterns in IG maps across different bands showing that the network has learnt to extract information about the photometry accordingly.
        However, there are only a certain amount of galaxies have been noticed by the network.
        This is possibly due to a bright star distracting the network from searching for galaxies.
        Therefore, a treatment for these bright stars such as incorporating their prior known coordinates would be a possible way to mitigate this bias.
        
        We observe that high impact features distribute at the central parts of the images, especially in \textit{gri}-bands.
        This could be related to the more noticeable contrast between the central region and the background, providing more information for the network.
        It is also encouraging to observe that bad columns and artefacts shown in \textit{riz}-bands are ignored by the extractor, which reduces the need to perform image pre-processing to remove these patterns.
        Additionally, we find that the network has extracted complicated features in fainter bands, such as \textit{u}-band shown in the first row of Fig.\ref{fig:ig}, demonstrating its potential of detecting faint objects in optical images.
        Based on the interpretation using IG, it is reassuring that in the classification task the network has learnt to extract meaningful features in the images.
        
    \subsection{Regression network}
    
\begin{figure*}
        \centering
        \includegraphics[width=2.\columnwidth]{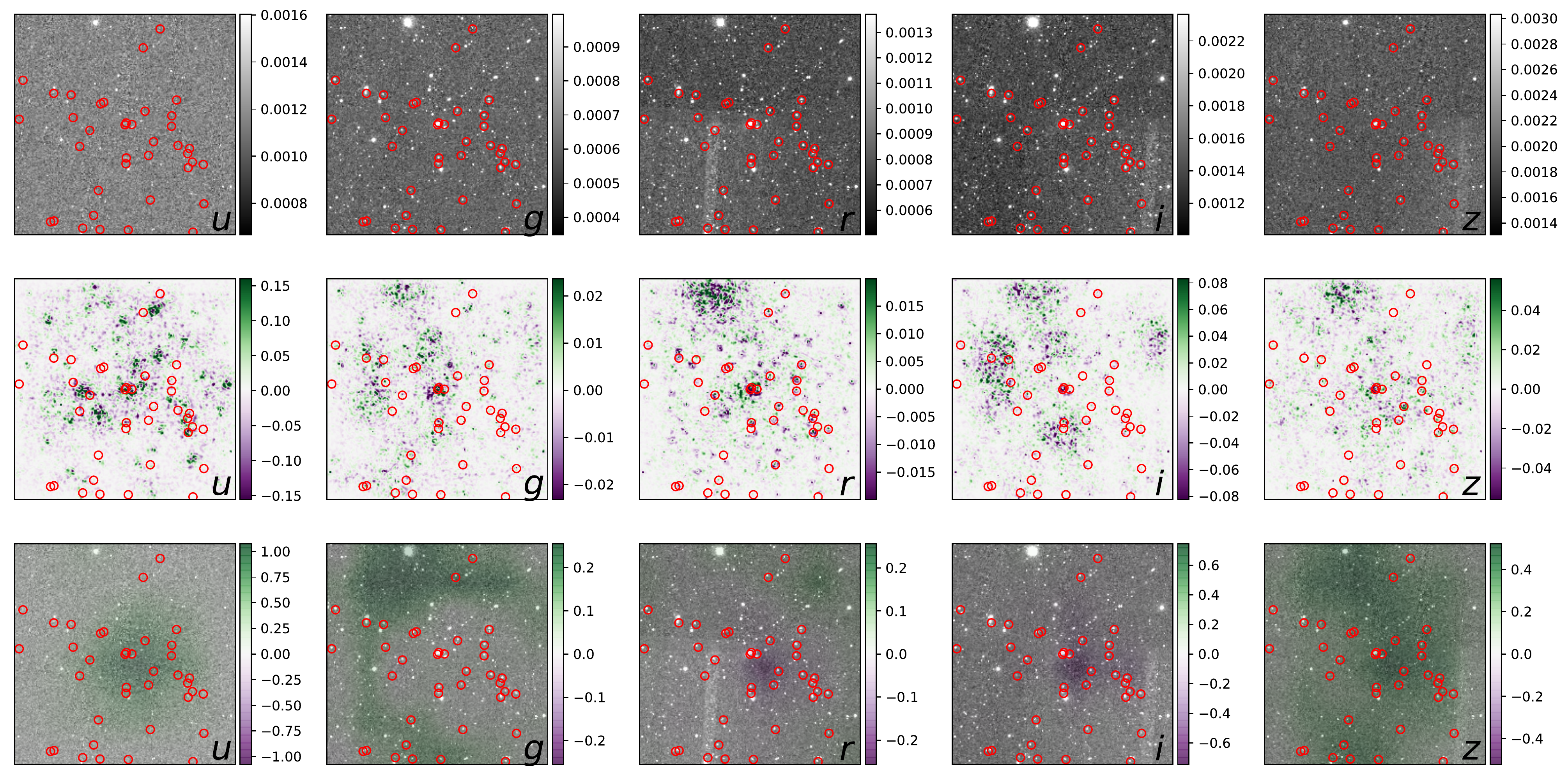}
        \caption{An example of the visualizations of the interpretations given by the IG and CAM methods across five bands.
                 The red open circles denote  the cluster members given by the redMaPPer algorithm.
                 The letter at the lower-right corner shows the photometric band of the image.
                 \textit{Upper panel}: the original image from SDSS as the input of the extractor.
                 \textit{Middle panel}: the resulting maps given by the IG applied on the extractor.
                 \textit{Lower panel}: the resulting maps given by the CAM+IG applied on the regression network, and the former is overlapped on the original image.}
        \label{fig:ig}
    \end{figure*}

        For the regression network, we combine IG with CAM given that the feature vectors have different dimensions from single-band 3-channel images.
        For details, we build the IG maps from the feature vectors with the corresponding regression outputs, de-project the maps back to the input images by treating the maps as the weights, and apply CAM on those weights.
        The resulting CAM+IG maps are shown in the bottom row of Fig.\ref{fig:ig}.
        
        The CAM+IG maps depict more abstract features compared to the patterns in the pure IG maps.
        This is because the inputs of the regression network are the products of higher layers of the feature extractor in which the spatial information has been compressed in the return of the increasing depth.
        The central regions in some of the bands but not all are highlighted. We do not find a significantly better performance for the cases with their central regions highlighted.
        This indicates that our network treats these features in a distinct way compared to the traditional algorithms that use the information of photometry and colour.
        Overall, we do find a consistent result from the visualizations of both methods, IG and CAM+IG. The impact of bright stars still presents in the regression training and no bad features or artifacts have been focused on by the regression network.
        The interpretations of both networks have touched on the direction of improving network abilities in the future.
    
\section{Discussion and summary}
\label{sec:summary}
    
    In this paper, we presented a transfer learning model consisting of two CNNs, the feature extractor and the regression network.
    We have utilized a semi-supervised approach to make full use of the entire cluster sample including those without ICM-based mass estimates.
    The combination of the transfer learning and the semi-supervised learning approaches overcomes the issue of lacking sufficient amount of labeled data.
    The improvement is confirmed by comparing the performance of our model with the plain ResNet-18 model.
    
    The first network captures the important features of the photometric images by learning to distinguish \textit{ugriz} bands.
    It correctly predicts photometric bands with an accuracy of $92\%$.
    The learnt features are then passed on to the second network which outputs the cluster masses.
    The mean absolute error of the predicted masses reaches $0.232$ dex.
    Cluster richness derived through the sophisticated catalogue processing adopted by redMaPPer is strongly correlated with ICM-based mass proxies.
    Our model is at the same level of accuracy as the redMaPPer.
    It is worth noting that our algorithm takes photometric images as the input directly while redMaPPer uses SDSS categorical data of galaxy.
    
    Most deep learning algorithms require a large amount of labeled data for training.
    In astronomy, this is mostly available from the computationally expensive simulations,
    It also takes heavy efforts to mimic real observations, including adding realistic astrophysical and instrumental backgrounds.
    Compared to other cluster mass estimates using deep learning algorithms, 
    our model overcomes the data hungry issue and can be trained directly on the real data. This work therefore comes closer to the real world.

    It is crucial yet challenging to explain the predictions made by deep learning models.
    We make use of the IG method to interpret the important features obtained by the extractor from the input images. We use the CAM method to de-project the IG maps from the regression network back to the inputs.
    The IG maps show that the galaxies in the images are important when determining the bands, especially those at the central regions of the field.
    However, we find that not all member galaxies are noticed by the feature extractor due to the bright stars appearing in the images.
    On the other hand, the CAM+IG method shows more abstract features compared to the IG maps.
    We observe that for the regression network, only certain bands have their central regions  highlighted, indicating a complicated process of handling the information of photometric bands is adopted by the network.
    It is worth noting that our algorithm has learnt to ignore most background, bad pixels, and artefacts, which potentially can be utilized to facilitate imaging processing and cataloguing.

    Here we point out a few directions to improve the model.
    Based on the model interpretation, we have shown that the bright stars have a noticeable impact on the feature extractor which undermines the performance.
    Therefore, certain treatment of bright stars is expected to improve the results.
    Deep learning results are known to be positively correlated with the size of the data set.
    Labeled data at the low mass end is particularly desirable in our case.
    Presumably, a more advanced model, such as ConvNeXt (\citealt{2022arXiv220103545L}), may give better results via extracting more representative features of galaxy clusters or handling the bright stars in a more creative way.
    Our model currently provides the point estimate of cluster mass without uncertainty measurements.
    It is crucial for future work to incorporate the uncertainty quantification into our model to provide robust estimation on cluster masses.
    A few techniques, including the variational inference via Bayesian neural network (\citealt{blundell2015weight}) and the ensemble learning (\citealt{lakshminarayanan2016simple}), have been proposed to quantify the uncertainties, and can be employed alongside our regression network.
    We stress that the performance of our model is likely to scale up in the near future given the in-flowing high resolution photometric images (e.g., DES, LSST) as well as the more accurate and consistent ICM-based mass estimates (e.g., eROSITA, CMB-S4), promised by the current and future multi-wavelength observations.

\section*{Acknowledgements}

We would thank the University of Kentucky Center for Computational Sciences and Information Technology Services Research Computing for their support and use of the Lipscomb Compute Cluster and associated research computing resources. 
S.-C.\ L. and Y.\ S. were supported by Chandra X-ray Observatory grant GO1-22126X, NASA grant 80NSSC21K0714, and NSF grant 2107711. 

Funding for the Sloan Digital Sky 
Survey has been provided by the 
Alfred P. Sloan Foundation, the U.S. 
Department of Energy Office of 
Science, and the Participating 
Institutions. 

SDSS acknowledges support and 
resources from the Center for High 
Performance Computing  at the 
University of Utah. The SDSS 
website is www.sdss.org.

SDSS is managed by the 
Astrophysical Research Consortium 
for the Participating Institutions 
of the SDSS Collaboration including 
the Brazilian Participation Group, 
the Carnegie Institution for Science, 
Carnegie Mellon University, Center for 
Astrophysics | Harvard \& 
Smithsonian, the Chilean Participation 
Group, the French Participation Group, 
Instituto de Astrof\'isica de 
Canarias, The Johns Hopkins 
University, Kavli Institute for the 
Physics and Mathematics of the 
Universe (IPMU) / University of 
Tokyo, the Korean Participation Group, 
Lawrence Berkeley National Laboratory, 
Leibniz Institut f\"ur Astrophysik 
Potsdam (AIP),  Max-Planck-Institut 
f\"ur Astronomie (MPIA Heidelberg), 
Max-Planck-Institut f\"ur 
Astrophysik (MPA Garching), 
Max-Planck-Institut f\"ur 
Extraterrestrische Physik (MPE), 
National Astronomical Observatories of 
China, New Mexico State University, 
New York University, University of 
Notre Dame, Observat\'ario 
Nacional / MCTI, The Ohio State 
University, Pennsylvania State 
University, Shanghai 
Astronomical Observatory, United 
Kingdom Participation Group, 
Universidad Nacional Aut\'onoma 
de M\'exico, University of Arizona, 
University of Colorado Boulder, 
University of Oxford, University of 
Portsmouth, University of Utah, 
University of Virginia, University 
of Washington, University of 
Wisconsin, Vanderbilt University, 
and Yale University.

\section*{Data Availability}

The data underlying this article will be shared on reasonable request to the corresponding author.



\bibliographystyle{mnras}
\bibliography{bibliography} 

\bsp	
\label{lastpage}
\end{document}